\documentclass{article}
\usepackage[preprint]{neurips_2021}
\usepackage[utf8]{inputenc} 
\usepackage[T1]{fontenc}    
\usepackage{hyperref}       
\usepackage{url}            
\usepackage{booktabs}       
\usepackage{amsfonts}       
\usepackage{nicefrac}       
\usepackage{microtype}      
\usepackage{xcolor}         
\usepackage{multicol}
\usepackage{multirow}
\usepackage{wrapfig}

\usepackage{amsmath,amsthm,mathtools,amsfonts,amssymb}
\usepackage{graphicx}
\usepackage{algorithm, algorithmic}
\usepackage{subfigure,subfloat}
\usepackage{soul}

\title{A prior regularized full waveform inversion using generative diffusion models}

\author{%
Fu Wang$^{1}$,~Xinquan Huang$^{1}$\thanks{Corresponding Author.}~,~Tariq Alkhalifah$^{1}$\\
$^1$King Abdullah University of Science and Technology\\
\texttt{\{fu.wang,xinquan.huang,tariq.alkhalifah\}@kaust.edu.sa}\\
}

\begin{document}
\maketitle

\title{A prior regularized full waveform inversion using generative diffusion models}

\begin{abstract}
Full waveform inversion (FWI) has the potential to provide high-resolution subsurface model estimations. However, due to limitations in observation, e.g., regional noise, limited shots or receivers, and band-limited data, it is hard to obtain the desired high-resolution model with FWI. To address this challenge, we propose a new paradigm for FWI regularized by generative diffusion models. Specifically, we pre-train a diffusion model in a fully unsupervised manner on a prior velocity model distribution that represents our expectations of the subsurface and then adapt it to the seismic observations by incorporating the FWI into the sampling process of the generative diffusion models. What makes diffusion models uniquely appropriate for such an implementation is that the generative process retains the form and dimensions of the velocity model.
Numerical examples demonstrate that our method can outperform the conventional FWI with only negligible additional computational cost. Even in cases of very sparse observations or observations with strong noise, the proposed method could still reconstruct a high-quality subsurface model. Thus, we can incorporate our prior expectations of the solutions in an efficient manner. We further test this approach on field data, which demonstrates the effectiveness of the proposed method.

\end{abstract}

\section{Introduction}
Full waveform inversion (FWI) plays a significant role in reconstructing high-resolution, subsurface models by iteratively matching the observed seismic records and the simulated data governed by the wave equation \citep{Tarantola1984}. 
However, limited by imperfect observations, this non-linear ill-posed problem can admit poor results at a high computational cost. 
To remedy this problem, a classical and powerful way is to use prior information as regularization to guide the inversion.
Standard regularization methods, including Tikhonov \citep{Aster2011} and total variation (TV) regularizations \citep{alkhalifah2018full}, often push the inversion to models that are either smooth or piecewise smooth, respectively.
Recently, sparsity-promoting (or constrained) FWI has aroused a lot of interest in the field of geophysical inversion problems. Curvelet or Seislet transforms can be used to conduct FWI under the framework of sparse model constraints to obtain robust inversion results \citep{Aster2011,Lix2012,Xue2017}. 
However, from the perspective of representation theory, the sparsity of the model is also highly dependent on the basis function, that is, the sparse representation of the model or data using the fixed basis functions cannot capture the special structural information in the model. 
In order to better grasp the characteristics of representation parameters, dictionary-based FWI \citep{7471912,Zhu2017,huang2019robust} methods were proposed to encode velocity models effectively and succinctly in a learning way. However, these methods are sensitive to the  values of the hyperparameters used.

With the recent popularity of deep learning algorithms in many fields, such as computer vision, speech recognition, and natural language processing \citep{lecun2015deep}, the trend in geophysical inversion is also levitating towards using DNN to learn a map directly between the data and model domains in a supervised way \citep{li2020dpfwi} or using prior model information obtained by deep learning methods from seismic velocity models, migration images or well information in regularizing FWI, explicitly \citep{li2021deep}. Compared to the latter, the network in the former method needs to be retrained using a sufficient amount of training pairs if the inference data are out of the distribution of the original training data \citep{alkhalifah2022mlreal}.
Alternatively, \cite{mosser2020stochastic} adopt a generative adversarial network (GAN) to represent the velocity prior distribution to regularize FWI, implicitly, and update the velocity model in latent space. However, GANs are difficult to train and their generative progress is not clear. On the other hand, diffusion models have shown state-of-the-art performance both as generative models and as unsupervised generative priors to solving linear and some nonlinear inverse problems by way of plug-and-play \citep{chung_improving_2022}. This can be attributed to their stable training and ability to track the progress of the generative process from coarse to fine details. Considering FWI is also an inverse problem often progressing from coarse to fine resolution, diffusion models should be suitable and helpful in regularizing FWI.

To this end, we propose a novel diffusion FWI, which embeds the prior velocity distribution represented by the diffusion model into FWI as a regularization term. Specifically, we pre-train a diffusion model on a prior distribution of the expected subsurface models, in a fully unsupervised manner and then adapt it to the seismic observations by incorporating the FWI into the sampling process of the generative diffusion models. By utilizing the background velocity information from well data or a good initial model in generating the training set for the diffusion model, that information will also act as a prior to the inversion process.
On the other side, considering the generative diversity of the diffusion model with different initializations and stochastic sampling, it also can be regarded as conditional generative progress guided by the gradient of the data fidelity term. 

To the best of our knowledge, our work is the first attempt to leverage diffusion models to enhance the performance of FWI. We showcase the effectiveness of the proposed generative diffusion-based FWI via numerical experiments, as well as field data. Specifically, the main contributions of this paper can be summarized as follows:
\begin{itemize}
    \item A novel diffusion FWI, which embeds the prior velocity distribution represented by the diffusion model into FWI.
    \item Our approach provides new insights into incorporating the prior information into FWI, and could be easily adapted to making use of multi-model prior and extended to 3D. 
    \item We evaluate the effectiveness of the proposed method on synthetic and field data and showcase its improvements in accuracy and resolution compared to conventional FWI.
\end{itemize}

The rest of the paper is organized as follows: we first introduce the diffusion model. To state our method clearly, we focus on acoustic FWI inversion regularized by diffusion models. Then we show the comparisons between conventional FWI and our method on synthetic data, considering situations in which we have noise, sparse observations, and observations lacking low-frequency components. We further demonstrate the effectiveness of the proposed method on field data. Finally, we share out thoughts and summarize our developments in the discussion and conclusion.

\section{Theory}
The proposed FWI using diffusion models includes two steps: pretraining a diffusion
model using random velocity models and implementing FWI regularized by diffusion models. 
\subsection{Generative diffusion models}
There are two important concepts in diffusion models: the forward data noising process (diffusion) and the inverse process. For the forward process, the noise is gradually added to corrupt the image until it becomes dominated by white Gaussian noise, while for the inverse process we train a neural network to reverse the process gradually until the image is retained. 
A simple and excellent variance-preserving diffusion model, denoising diffusion probabilistic models (DDPMs), have shown promising performance in image synthesis. \cite{song2020score} unified it into a score-based generative modeling approach and defined the forward noising process of the data, from the perspective of score matching, with a linear stochastic differential equation 
\begin{equation}
d \boldsymbol{x}=-\frac{\beta(t)}{2} \boldsymbol{x} d t+\sqrt{\beta(t)} d \boldsymbol{w}, t\in[0,1],
\end{equation}
where $\boldsymbol{x}(t=0)$ (omitted as $\boldsymbol{x}_0)$ are the samples from the data distribution, 
$\beta(t)$ denotes the noise schedule to make sure the final corrupted data $\boldsymbol{x}_1$ satisfies the pre-defined noise distribution (standard Gaussian distribution), and $\boldsymbol{w}$ is the standard Wiener process (Brownian motion). The $d$ in front of these parameters describes the change in them as a function of the noising iteration (time). 
We set the prior distribution of $\boldsymbol{x}$ as $p(\boldsymbol{x})$ and $\boldsymbol{x}_t\sim p_t(\boldsymbol{x})$ for time horizon $t\in[0,1]$. 
Then the reverse process is given by
\begin{equation}
\label{equ:r-diff}
d \boldsymbol{x}=\left[-\frac{\beta(t)}{2} \boldsymbol{x}-\beta(t) \nabla_{\boldsymbol{x}} \log p_t\left(\boldsymbol{x}\right)\right] d t+\sqrt{\beta(t)} d \overline{\boldsymbol{w}},
\end{equation}
where $\nabla$ denotes the gradient operation, $\nabla_{\boldsymbol{x}} \log p_t\left(\boldsymbol{x}\right)$ is the score function \cite{liu2016kernelized} of distribution $p_t(\boldsymbol{x})$, 
and the $\overline{\boldsymbol{w}}$ denotes a standard Wiener process in the reverse-time direction \citep{song2020score}. 
\begin{figure}[!htb]
  \centering
  \includegraphics[width=0.95\columnwidth]{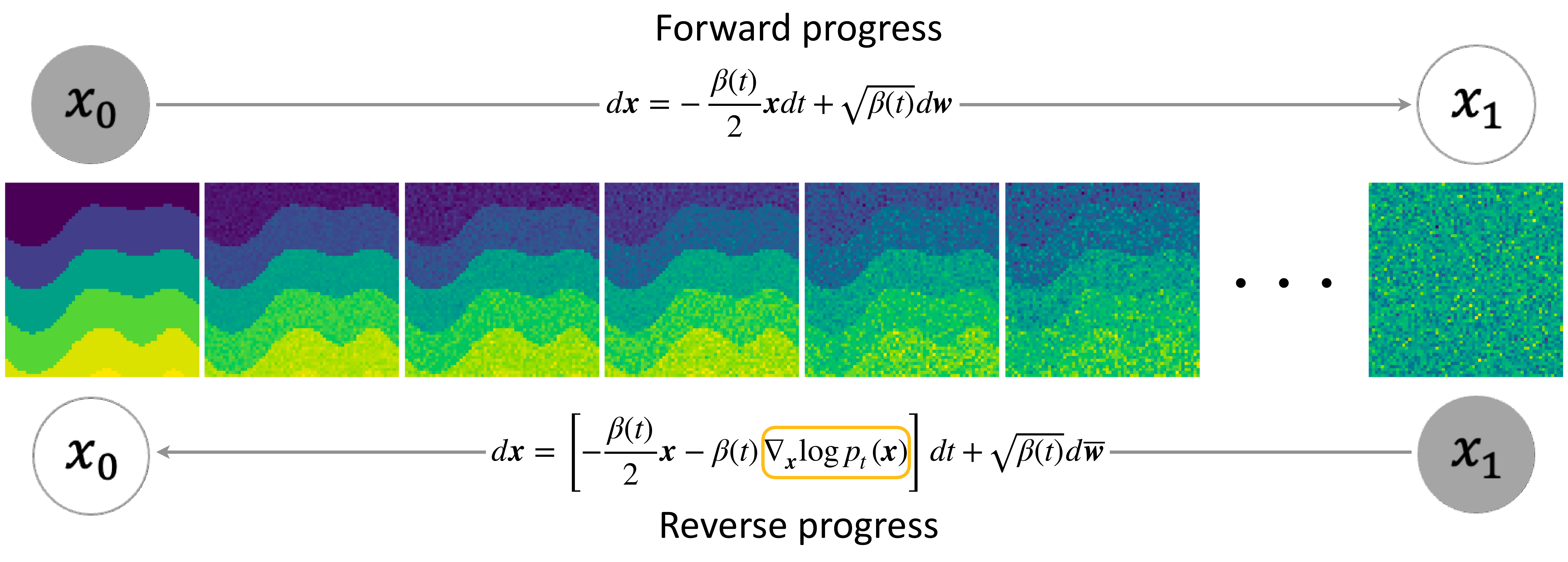}
  \caption{The workflow of the diffusion process and its reverse process.}
\label{Forward_reverse}
\end{figure}
The workflow is shown in Figure~\ref{Forward_reverse}.
The key point here is to learn this score function via a neural network with the parameter $\boldsymbol{\theta}$.
In practice, we use the denoising score matching to approximate it:
\begin{equation}
\begin{array}{r}
\boldsymbol{\theta}^*=\underset{\boldsymbol{\theta}}{\arg \min } \mathbb{E}_t\left\{\mathbb{E}_{\boldsymbol{x}_0}\mathbb{E}_{\boldsymbol{x}_t|\boldsymbol{x}_0} \left[\| \mathbf{s}_{\boldsymbol{\theta}}(\boldsymbol{x}_t, t)\right.\right. \\
\left.\left.-\nabla_{\boldsymbol{x}_t} \log p_{0 t}(\boldsymbol{x}_t \mid \boldsymbol{x}_0) \|_2^2\right]\right\},
\end{array}
\end{equation}
where $p_{0 t}(\boldsymbol{x}_t \mid \boldsymbol{x}_0)$ denotes the transition distribution from $\boldsymbol{x}_0$ to $\boldsymbol{x}_t$ and $\boldsymbol{x}_t$ is sampled from $p_{0 t}(\boldsymbol{x}_t \mid \boldsymbol{x}_0)$. 
The theory behind denoising score matching ensures that $\boldsymbol{s}_{\boldsymbol{\theta}^*}\left(\boldsymbol{x}_t, t\right)$ can approximate $\nabla_{\boldsymbol{x}} \log p_t\left(\boldsymbol{x}\right)$ well \citep{song2020score}.

\begin{figure}[!htb]
  \centering
  \includegraphics[width=0.95\columnwidth]{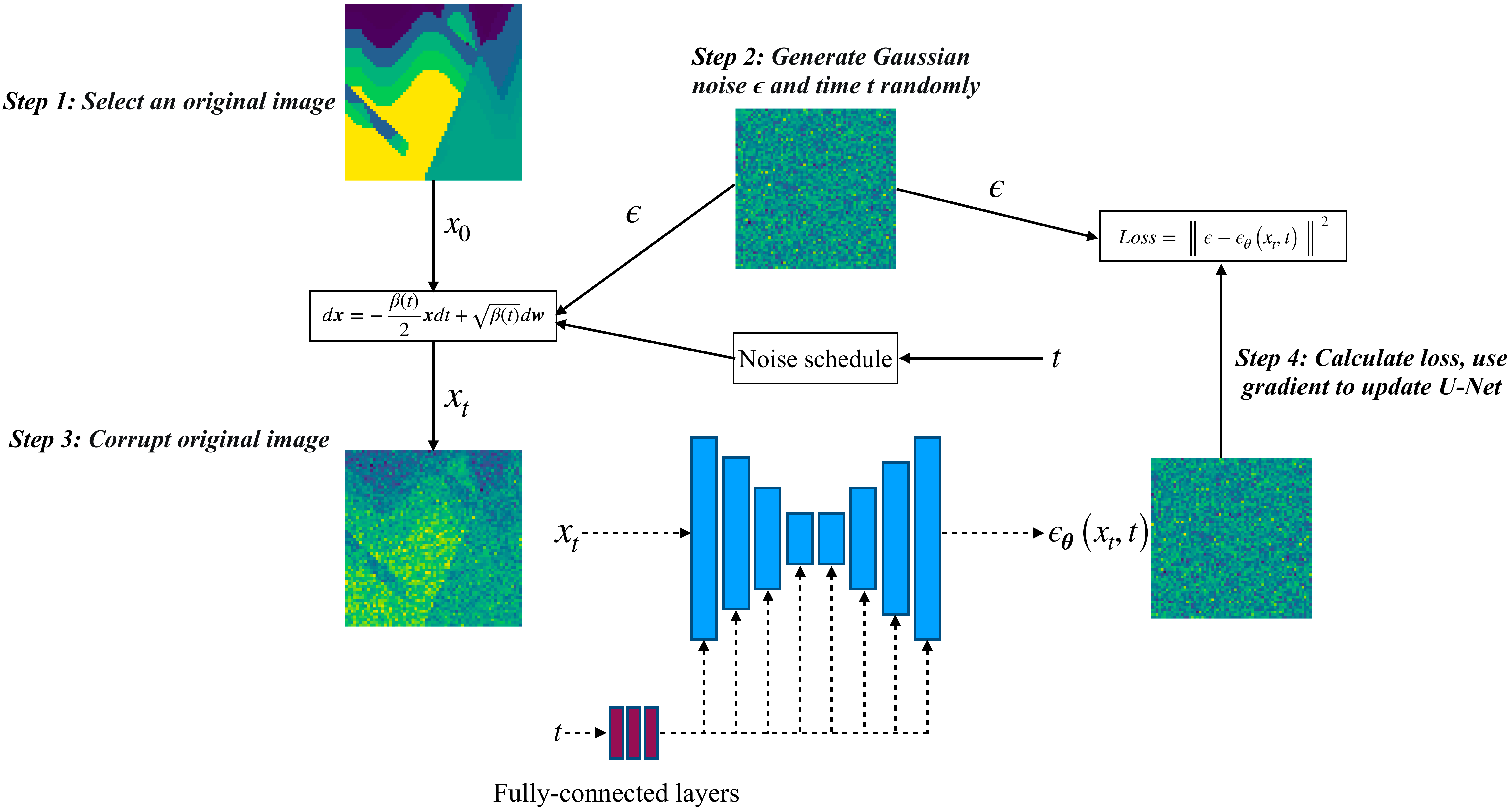}
  \caption{The training pipeline of the diffusion models.}
  \label{fig:training-ddpm}
\end{figure}

The beauty of this approach is that once the neural network is trained, we can use $\boldsymbol{s}_{\boldsymbol{\theta}^*}\left(\boldsymbol{x}_t, t\right)$ in equation~\ref{equ:r-diff}, where we obtain 
\begin{equation}
d \boldsymbol{x}=\left[-\frac{\beta(t)}{2} \boldsymbol{x}-\beta(t) \boldsymbol{s}_{\boldsymbol{\theta}^*}\left(\boldsymbol{x}_t, t\right)\right] d t+\sqrt{\beta(t)} d \overline{\boldsymbol{w}}=f(\boldsymbol{x},\boldsymbol{s}_{\boldsymbol{\theta}^*},t)+\sqrt{\beta(t)} d \overline{\boldsymbol{w}},
\end{equation}
to sample $\boldsymbol{x}_0$ from the prior distribution $p(\boldsymbol{x}_0)$ and then use it in inverse problems as a prior. 
Specifically, as shown in Figure~\ref{fig:training-ddpm}, the training progress of diffusion models includes the following steps:
\begin{itemize}
    \item Select a batch of random training samples;
    \item Generate Gaussian noise to corrupt the clean samples (velocity models), which is referred to as the diffusion process over time $t$;
    \item Feed the noisy samples for different $t$ into a U-Net to output an estimation of the added noise $\epsilon_\theta\left(x_t, t\right)$;  
    \item Measure the loss by taking the difference between the estimated and the reference noise and then back propagate the difference to update the U-Net. 
\end{itemize}
This training procedure of the Unet is performed over training samples and noising iterations (time), and thus, can be costly. 
However, we only need to train once for a specific velocity distribution, and thus, the training process can be considered as an overhead cost.
We will discuss this matter in detail in the pretraining of the diffusion model section.

\subsection{Diffusion regularized FWI}
\begin{figure*}[!htb]
  \centering
  \includegraphics[width=0.85\textwidth]{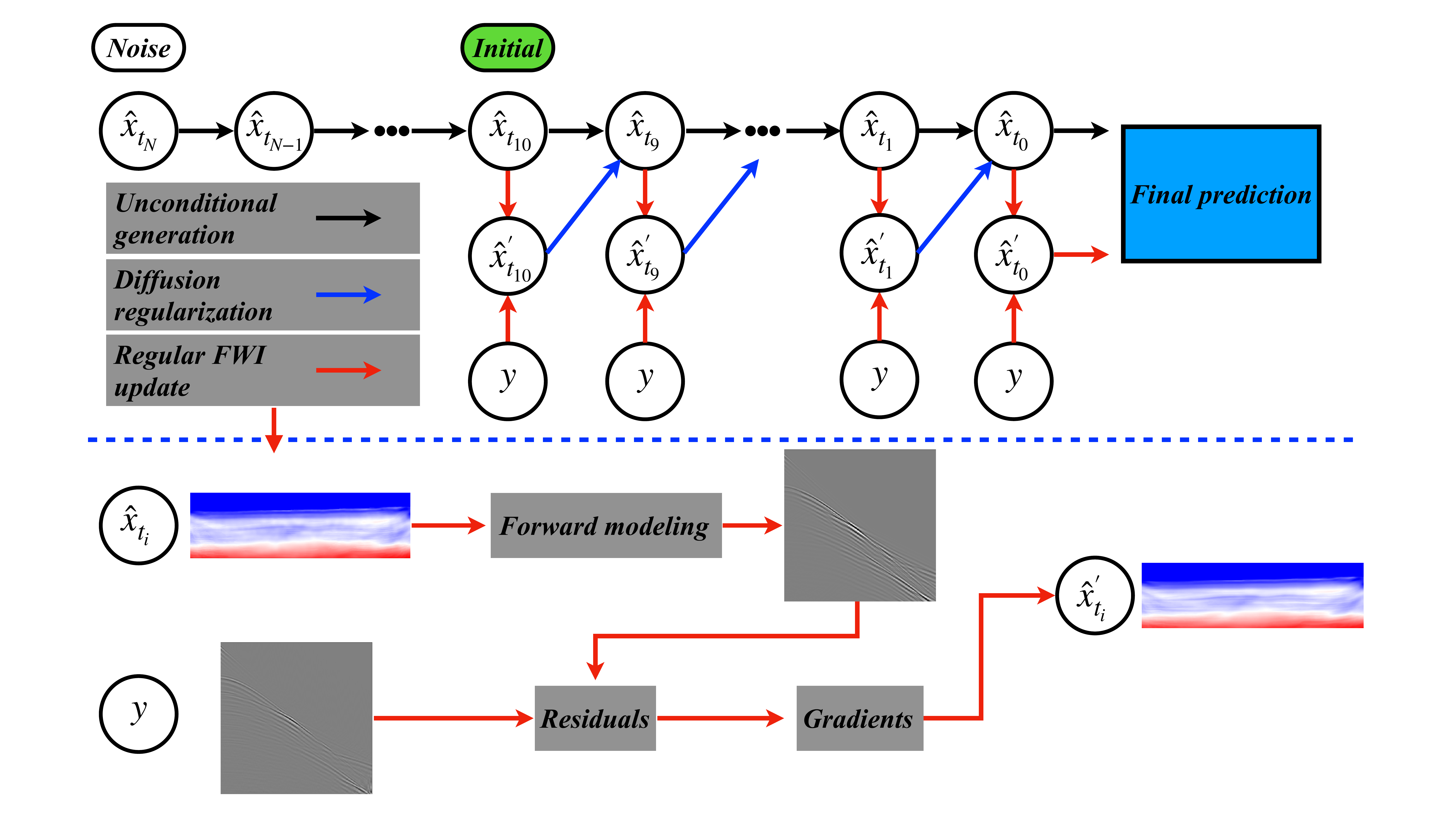}
  \caption{(Top) An overview of our method for the generative diffusion model-based FWI. The black arrow represents the original unconditional inverse diffusion progress. The red and blue arrows jointly represent the workflow of our diffusion FWI, where the red one represents a regular FWI iteration, and the blue one denotes the inverse diffusion progress. Starting from a certain reverse diffusion time step and an initial velocity model, we do a few inner iterations of conventional FWI followed by the velocity update using the diffusion model. (Bottom) An illustration of how to update FWI using the current velocity model and observed data. This also can be regarded as using the seismic observed data to guide the generation of the velocity model from the diffusion process.}
\label{diffusionFWI}
\end{figure*}

The objective function $J$ of a classical FWI with a regularization term is described as follows:
\begin{equation}
\label{equ:fwi}
J=\frac{1}{2}\left\|\mathbf{d}_{\mathrm{obs}}-\mathbf{F}(\mathbf{m})\right\|_2^2 + \mathbf{R(m)},
\end{equation}
where $\mathbf{d}_{obs}$ is the observed data, $\mathbf{m}$ is the velocity model, $\mathbf{F}$ is the wave equation operator (like that for the acoustic wave equation with constant density), and $\mathbf{R}$ denotes an implicit or explicit regularization term on the velocity model. We can use the proximal operator of $\mathbf{R}$, $\operatorname{prox}_{\mathbf{R}}$, to solve equation~\ref{equ:fwi}, 
\begin{equation}
\begin{split}
\mathbf{m}_{k+1}&=\operatorname{prox}_{\mathbf{R}}\left(\mathbf{m}_k+\alpha_k \frac{\partial J}{\partial \mathbf{m}_k}\right) \\ 
    &=f\left(\mathbf{m}_k+\alpha_k \frac{\partial J}{\partial \mathbf{m}_k}, \boldsymbol{s}_{\theta^*}, t\right) +\sqrt{\beta(t)} d \overline{\boldsymbol{w}},
\end{split}
\label{equ:proxi}
\end{equation}
where $\mathbf{m}_k$ is the velocity model at the kth iteration and $\alpha_k$ is the step length. The proximal operator has the form of a Gaussian denoiser in a Maximum-a-posterior (MAP) sense. Inspired by this fact, we choose $f$ corresponding to a reverse diffusion step to replace it, then we obtain the right-hand side term in equation~\ref{equ:proxi}.
Unlike the traditional iterative FWI update per iteration, here we incorporate the FWI process into the reverse progress (sampling) of the diffusion model, which we call diffusion regularized FWI. 
The diagram of the diffusion regularized FWI is shown in Figure~\ref{diffusionFWI}.
For every diffusion time step $t$, we do a few inner iterations of conventional FWI (could be 1, but in our case, we choose 10) followed by the velocity update using equation $f$. 
Unlike in the case of unconditional sampling, here we start from an initial velocity model for FWI.
\section{The pretraining of denoising diffusion probabilistic models}
The pretraining of diffusion models is essential in the proposed method as it is used to store the distribution of the subsurface velocity models and provide prior information for FWI. 
In this section, we will describe the parameters used to train the diffusion model based on the training steps described earlier.
Here, we train the diffusion model on parts of the OpenFWI dataset \citep{deng2021openfwi} to learn the distribution of these subsurface velocities as prior for the FWI. 

\begin{figure}[!htb]
  \centering
  \includegraphics[width=0.95\textwidth]{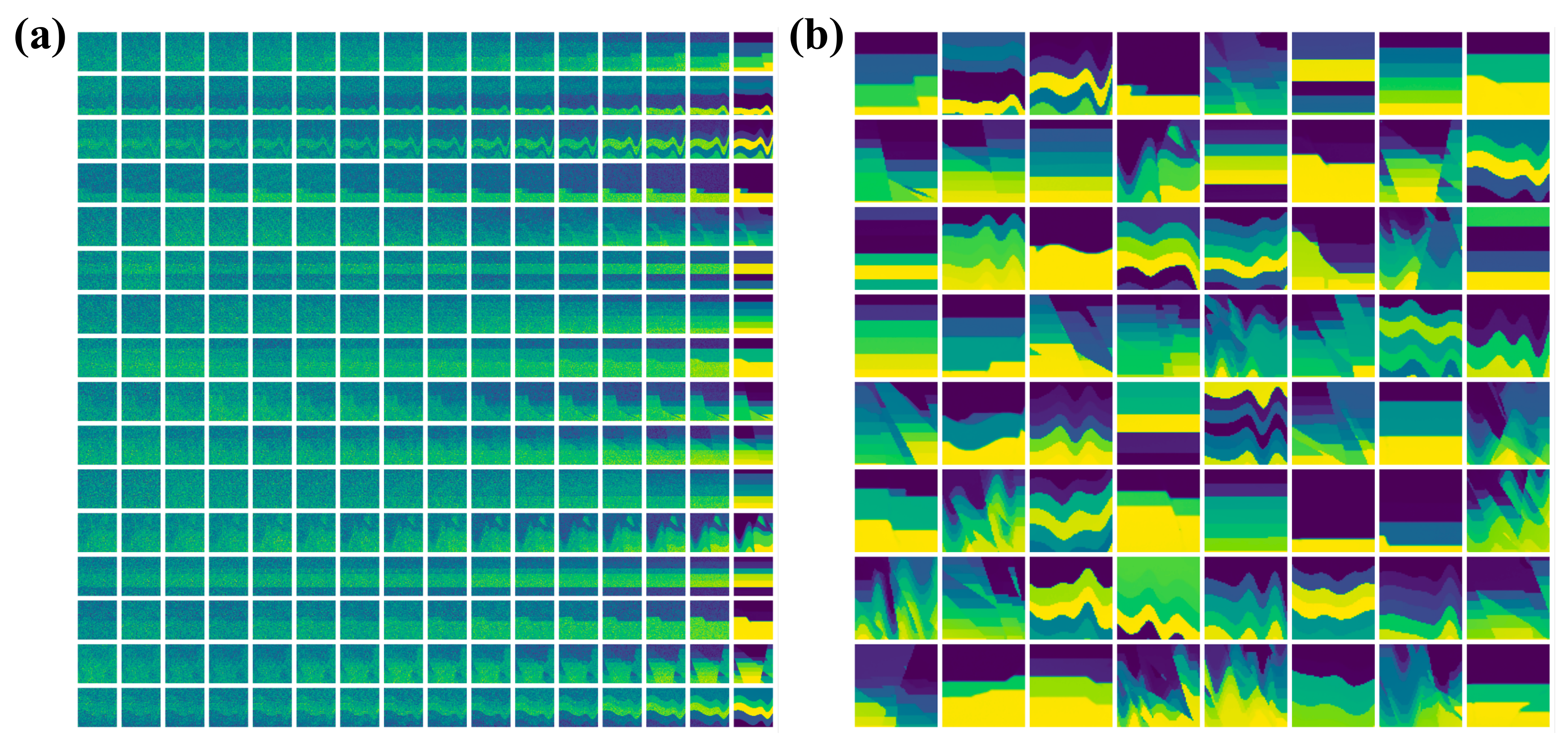}
  \caption{The generative progress a) from left to right for 16 velocity model samples (rows) using the diffusion model; b) shows 64 generated velocity models.}
  \label{generative_progress_results}
\end{figure}

We set the maximum timestep of the diffusion process to 500 and use a constantly increasing linear scheduler (forward process variance) for noise schedule $\beta_t$, where the starting $\beta_0=10^{-4}$ and the final $\beta_1=0.02$. 
As for the dataset, we extract 84000 velocity samples for training while testing on the other unseen velocities, and the size of the velocity is cropped to 64$\times64$. 
The backbone of the proposed method is a UNet model with attention and timestep embedding. 
Our neural network model uses four feature map
resolutions (64$\times$64 to 16$\times$8) and includes two convolutional residual blocks.
The diffusion timestep $t$ is added via the sinusoidal positional encoding into each residual block. 
We train the model for 100 epochs using an Adam optimizer with a learning rate of 5e$^{-4}$. Figure~\ref{generative_progress_results}a shows the generation progress of velocity models, where each column denotes the generated velocity along the timestep of diffusion models (visualized every 8 timesteps from timestep 120 to 0, from a total of 1000 steps), and each row denotes a different trajectory of sampling progress (targets to different generated models). 
The generation of the velocity model is equal to the reverse process for the diffusion on the true velocity model.
Figure~\ref{generative_progress_results}b represents some of the final generated velocity models drawn from the same distribution as the training velocity models. We observe that various velocity models can be sampled from the distribution stored in the diffusion model. 
Then with this pretrained diffusion model, we can combine it with FWI to provide more accurate and high-resolution inversion results.

\section{Numerical Experiments}
\begin{figure}[!htb]
  \centering
  \includegraphics[width=0.66\columnwidth]{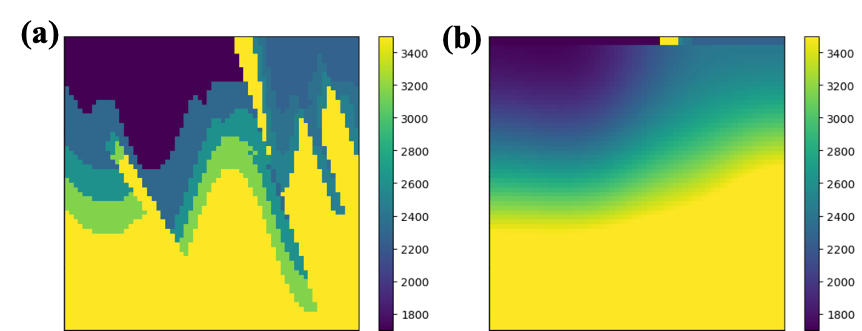}
  \caption{The true velocity model a) and the initial velocity model b).}
\label{true_model}
\end{figure}

After the pertaining of diffusion models, given seismic data, and a wavefield modeling engine, we use the proposed sampling method to reconstruct the subsurface velocity to demonstrate the effectiveness of our method. As we mentioned before, we can directly apply this pre-trained model in FWI without any retraining or adjustment. 
We use the velocity model (Figure~\ref{true_model}a) with a resolution of 64$\times$64 and a spatial interval of 10 m in both $x$ and $z$ directions, from the same velocity distribution, but not included in the training set, to demonstrate the performance of our proposed method for different acquisition scenarios, including coarse recording and the presence of noise. 
We generated data by numerically solving the acoustic wave equation with a time sampling interval of 1 ms using a 15 Hz Ricker wavelet with a total length of recording of 1.5 s. 

\begin{figure}[!htb]
  \centering
  \includegraphics[width=0.66\columnwidth]{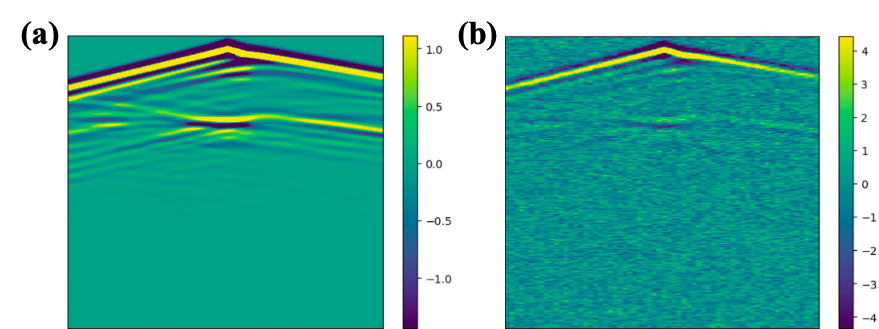}
  \caption{Simulated synthetic shot gathers recorded at 320 m a), and the corresponding noisy data b).}
\label{fig:syn_record}
\end{figure}

\begin{figure}[!htb]
  \centering
  \includegraphics[width=0.66\columnwidth]{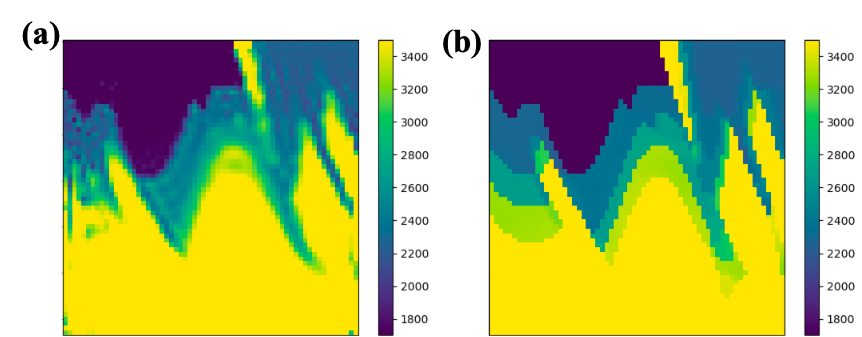}
  \caption{A comparison between the conventional FWI result (a) and the diffusion FWI result (b) for the clean synthetic data, with a sample shot gather shown in Figure \ref{fig:syn_record}a.}
\label{fig:fwi_clean}
\end{figure}

We uniformly place 64 receivers and 32 shots on the surface. 
We smooth the true velocity model to obtain the initial velocity for FWI (Figure~\ref{true_model}b).
Figures~\ref{fig:syn_record}a and \ref{fig:syn_record}b are the clean and noisy data, respectively. 
We first test the proposed and conventional FWI on clean data, with results shown in Figure~\ref{fig:fwi_clean}. Unlike the conventional FWI, whose resolution and accuracy are limited by the main frequency of the data, the diffusion regularized FWI provides high-resolution results courtesy of the prior. The details of the velocity model are reconstructed, which are hard to obtain using conventional methods.
That means our method can generate a velocity model, which is consistent with both the prior distribution and measurements. 

\begin{figure}[!htb]
  \centering
  \includegraphics[width=0.66\columnwidth]{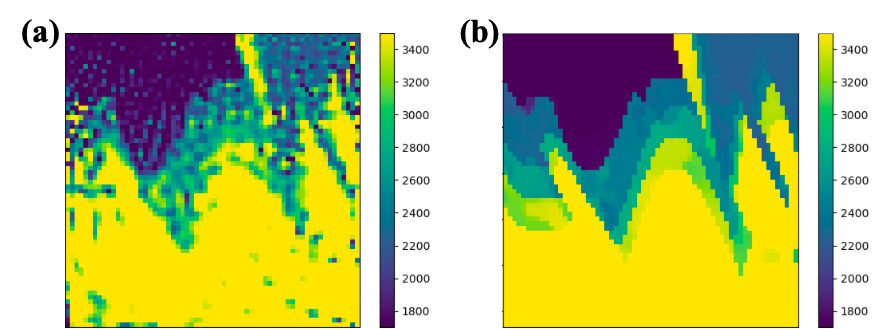}
  \caption{A comparison between the conventional FWI result (a) and the diffusion FWI result (b) for the noisy data, with a sample shot gather shown in Figure \ref{fig:syn_record}b.}
\label{fig:fwi_noise}
\end{figure}

As for the robustness against noise, as shown in Figure~\ref{fig:fwi_noise}, the conventional FWI fails to converge to an adequate solution and the inversion results are contaminated with noise while the proposed method still inverts an accurate and high-resolution velocity model.
The prior information stored in the diffusion model helped guide FWI to a clean and more accurate result.

\begin{figure}[!htb]
  \centering
  \includegraphics[width=1.0\columnwidth]{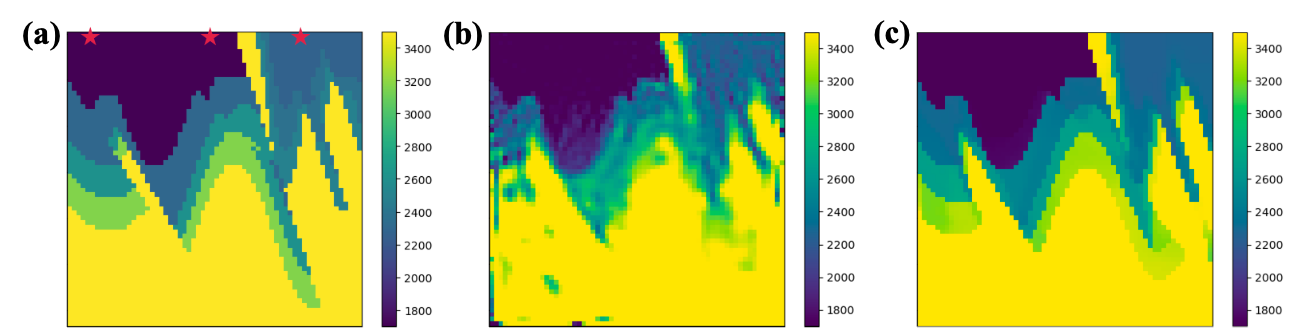}
  \caption{A comparison between the conventional FWI b),  and diffusion FWI c). a) is the true velocity model where the red stars are the source locations of the 3 shots, and b) and c) are the corresponding results for the 3 shot gathers. }
\label{fig:3shots}
\end{figure}

Besides noise recorded in shot gathers, it is common that we have only limited shots in field data, which results in insufficient illumination and influences the inversion results.
Thus, we test our approach for the case of sparse acquisition, where we only use three shots (Figure~\ref{fig:3shots}a). 
The inversion results are shown in Figures~\ref{fig:3shots}b and c. 
Due to the limited recorded data, the inversion of the conventional FWI contains a lot of errors in the velocity estimation. In contrast, thanks to the guidance of the prior information stored in the diffusion model, the inversion result using the proposed method is quite good, and high-resolution details are still recovered. 
A few accurate measurements could guide the diffusion model to sample the correct velocity, which is consistent with both the data and the prior distribution. 

\begin{figure}[!htb]
  \centering
  \includegraphics[width=1\columnwidth]{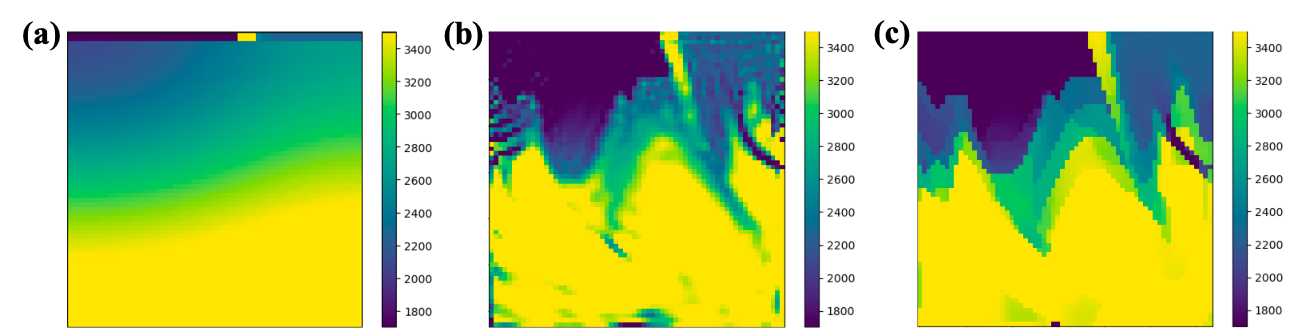}
  \caption{A comparison between the conventional FWI b),  and diffusion FWI c). a) is the initial model, which is much smoother than that in Figure~\ref{true_model}b, and b) and c) are the corresponding results using conventional FWI and diffusion regularized FWI in case of data without low frequencies. }
\label{fig:low_frequency}
\end{figure}

Finally, to further test the proposed method's potential for field applications, we show the case where the data lack low frequencies. 
Here, we smooth the initial model (Figure~\ref{true_model}b) further to test on a more challenging initial model (Figure~\ref{fig:low_frequency}a). 
We applied a highpass filter to remove recorded data with frequencies under 5 Hz and then apply the FWIs.
The results are shown in Figure~\ref{fig:low_frequency}.
Even though the result of our method (Figure~\ref{fig:low_frequency}b) is not as good as what we had before, it still has obvious improvements over the one from the conventional FWI (Figure~\ref{fig:low_frequency}c).

The above experiments on synthetic examples demonstrate the effectiveness of the proposed method in terms of enhancing the accuracy and resolution of the inverted model and also show the potential of the proposed method for field applications. Thus, we further test the approach on field data.

\section{Field data application}
\label{sec:fielddata}
\begin{figure}[!htb]
  \centering
  \includegraphics[width=0.35\columnwidth]{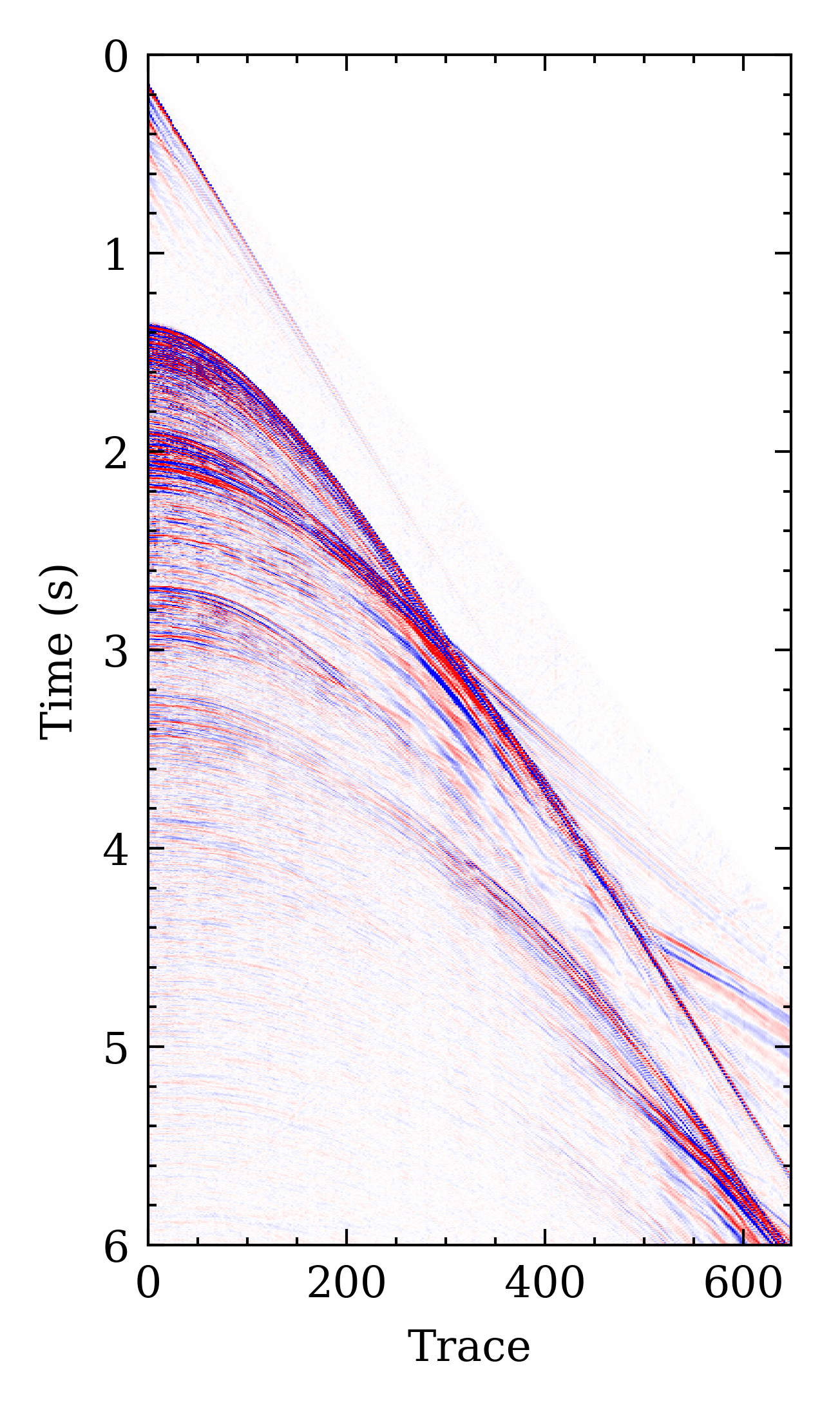}
  \caption{A raw shot gather from the field data for a source at a distance of 4.5 km.}
  \label{fig:field_data_record}
\end{figure}

Here, we use a 2D field marine dataset from North-Western Australia Continental
shelf acquired by CGG with a variable depth streamer to validate our method. 
Among the original 1824 shots with an 18.75 m horizontal interval between them, we chose 116 shots with a 98.75 m interval to test the inversion. Each shot gather includes 648 receivers at a 12.5 m spacing interval, and the corresponding minimum and maximum offsets are 16.9 m and 8.256 km, respectively. The maximum recording time is about 7 s with a 2 ms sampling interval. We use only 6 s of the recorded time, as shown in Figure~\ref{fig:field_data_record}.
The velocity model is designed to be 12.5 km long and 3.7 km deep, both with 12.5 m grid intervals.

\begin{figure}[!htb]
  \centering
  \includegraphics[width=0.85\textwidth]{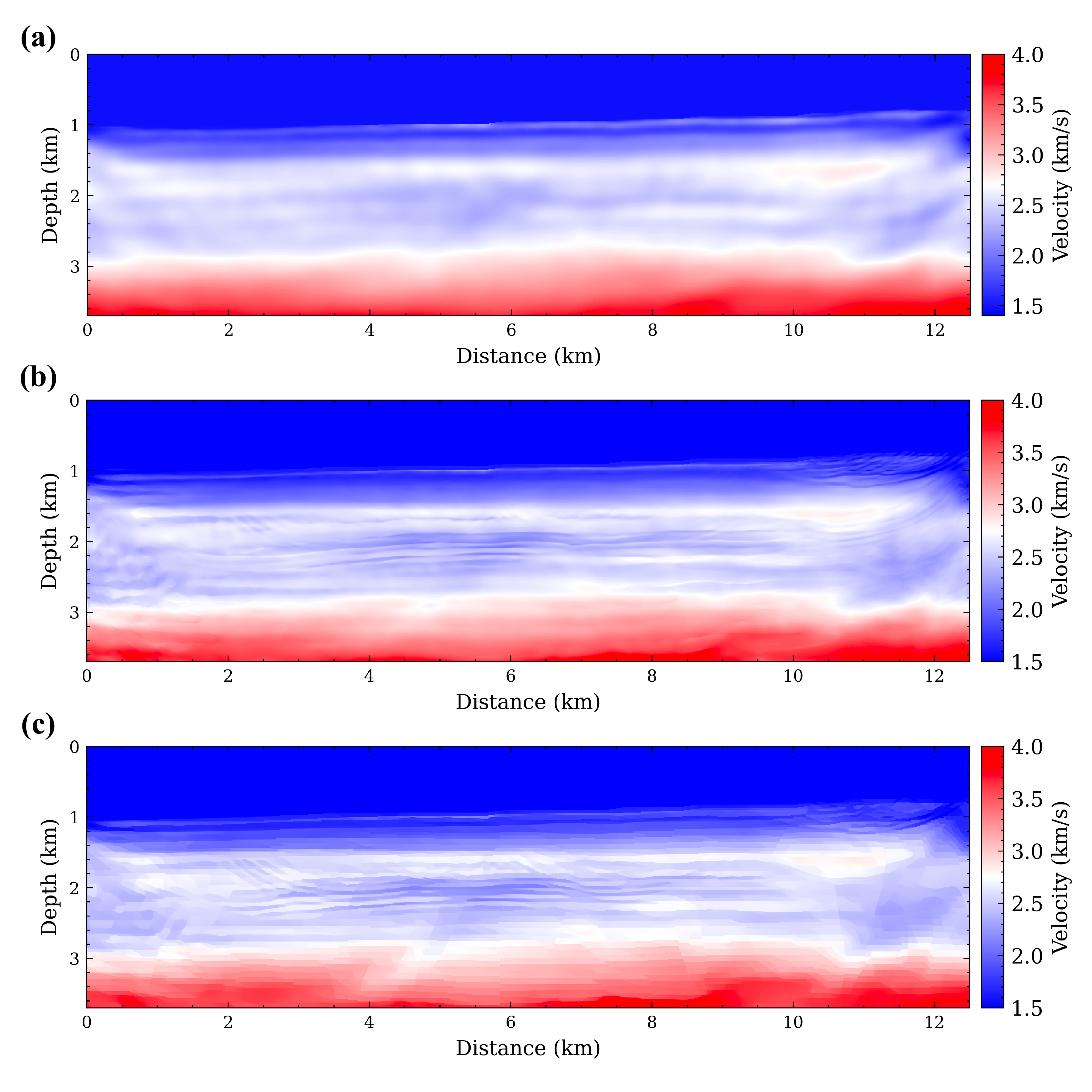}
  \caption{The inversion results of the field marine dataset. (a) The initial velocity, (b) the inverted velocity with the conventional FWI, (c) the inverted velocity using our diffusion regularized FWI.}
\label{fieldresult}
\end{figure}

We first applied a low pass filter on the field data to maintain a maximum frequency of 8.0 Hz. 
We estimate the source wavelet by inverting the near-offset early arrival seismograms with a constant water velocity of 1500 m/s and then use the following sequences of processing steps, including a brute-stack time-domain velocity analysis \citep{kalita2019flux} followed by an extended full waveform inversion with matching filter \citep{li2021extended}, to obtain a good initial model (shown in Figure~\ref{fieldresult}a).

For conventional FWI, we applied time-domain full waveform inversion using DeepWave package \citep{richardson_alan_2022}. We use an Adam optimizer with a learning rate of 0.003 and executed 100 iterations. Then we applied the proposed FWI using the previous pre-trained model based on the OpenFWI velocity models. Due to the poor performance of high dimensional diffusion models, we do not directly use a resolution of 296$\times$1000. Instead, we generated model patches with a resolution of 152$\times$152 and then merged them by means of a sliding window into the original velocity model size. 

\begin{figure}[!htb]
  \centering
  \includegraphics[width=0.75\textwidth]{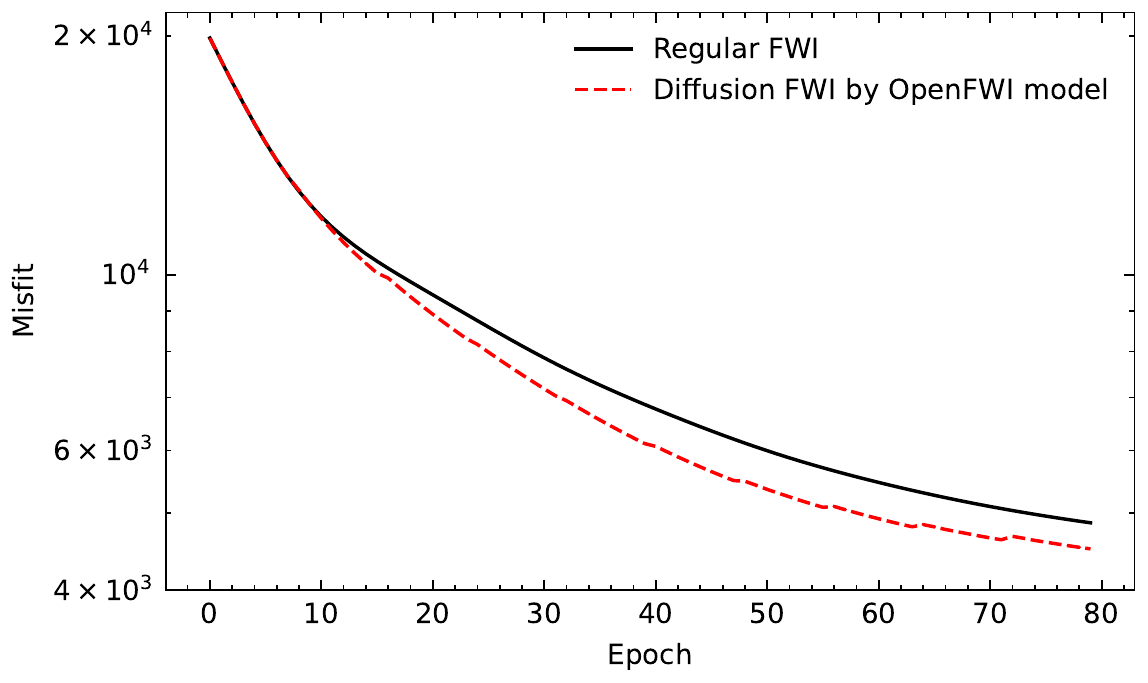}
  \caption{Convergence histories of the conventional FWI (black line) and the proposed diffusion regularized FWI (red dash line).}
\label{field_loss}
\end{figure}

The comparison of convergence histories (data misfit) between the conventional and diffusion FWI methods is shown in Figure~\ref{field_loss}, and we found that the proposed method converges faster and yields lower residuals. With the increasing impact of the diffusion models, the data residuals decrease faster. It implies that in spite of using regularization, it managed to fit the data better, which reflects the positive impact of our prior knowledge on the inversion. 

\begin{figure}[!htb]
  \centering
  \includegraphics[width=0.8\columnwidth]{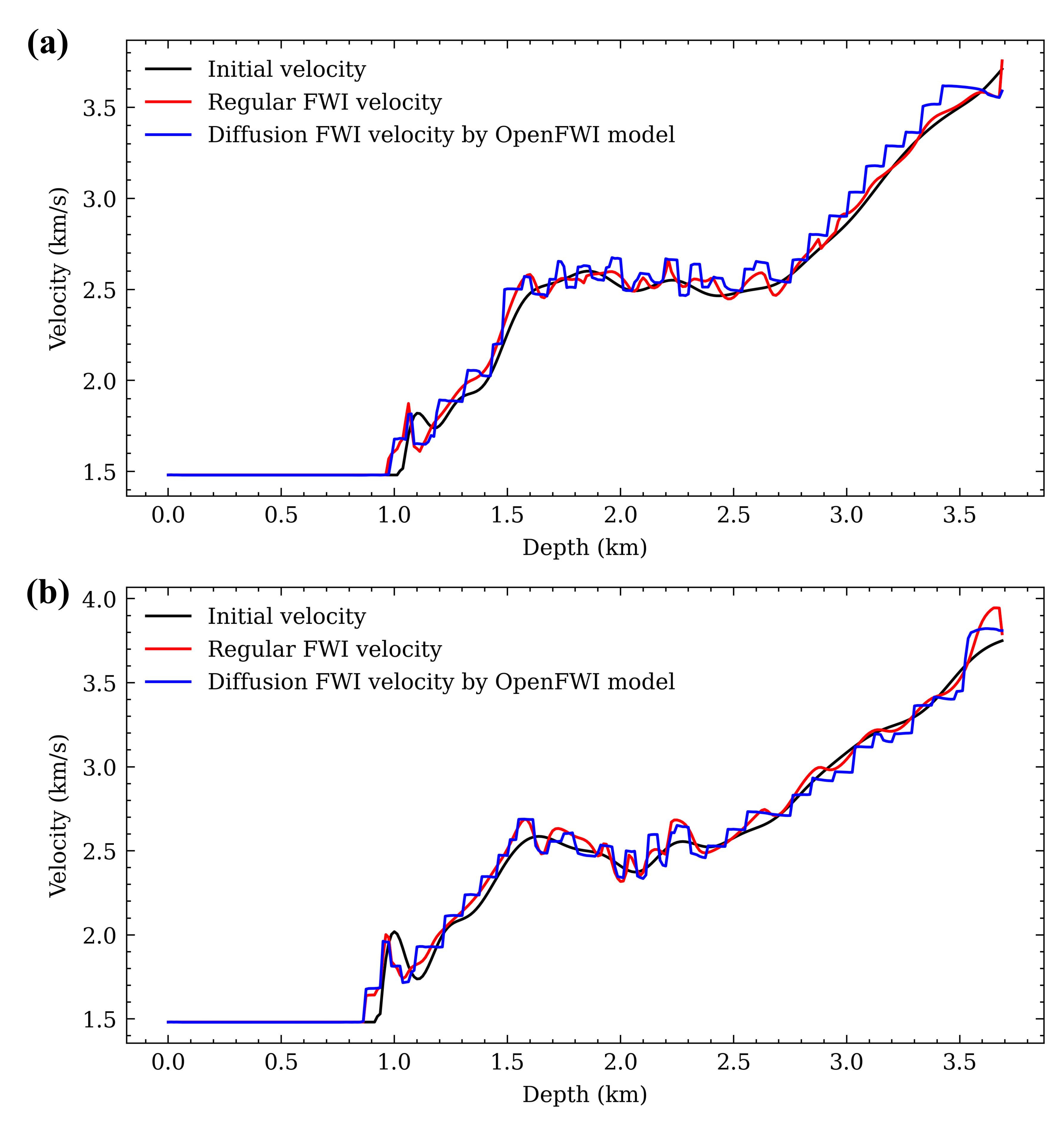}
  \caption{A comparison between vertical profiles at the lateral locations of 2.5 km (a) and 7.5 km (b), where the black lines denote initial velocity, the red lines denote the inverted result using conventional FWI, the blue lines denote that using the diffusion regularized FWI based on the pre-trained model using OpenFWI velocity models.}
  \label{profile}
\end{figure}

The inverted velocities are shown in Figure \ref{fieldresult}, and we observe that with the storage of the high-resolution information from the OpenFWI velocity models in our diffusion model, we can obtain a higher-resolution inverted velocity model with higher wavenumber content in the vertical direction compared to the result of conventional FWI. We extract the vertical profiles at the lateral locations of 2.5 km and 7.5 km and show them in Figure~\ref{profile}. Although the trend of the inverted velocity of the conventional FWI is consistent with that using the proposed method, the diffusion regularized FWI provides higher-wavenumber content and more details. This demonstrates the effectiveness of the proposed method.

\begin{figure}[!htb]
  \centering
  \includegraphics[width=1.0\textwidth]{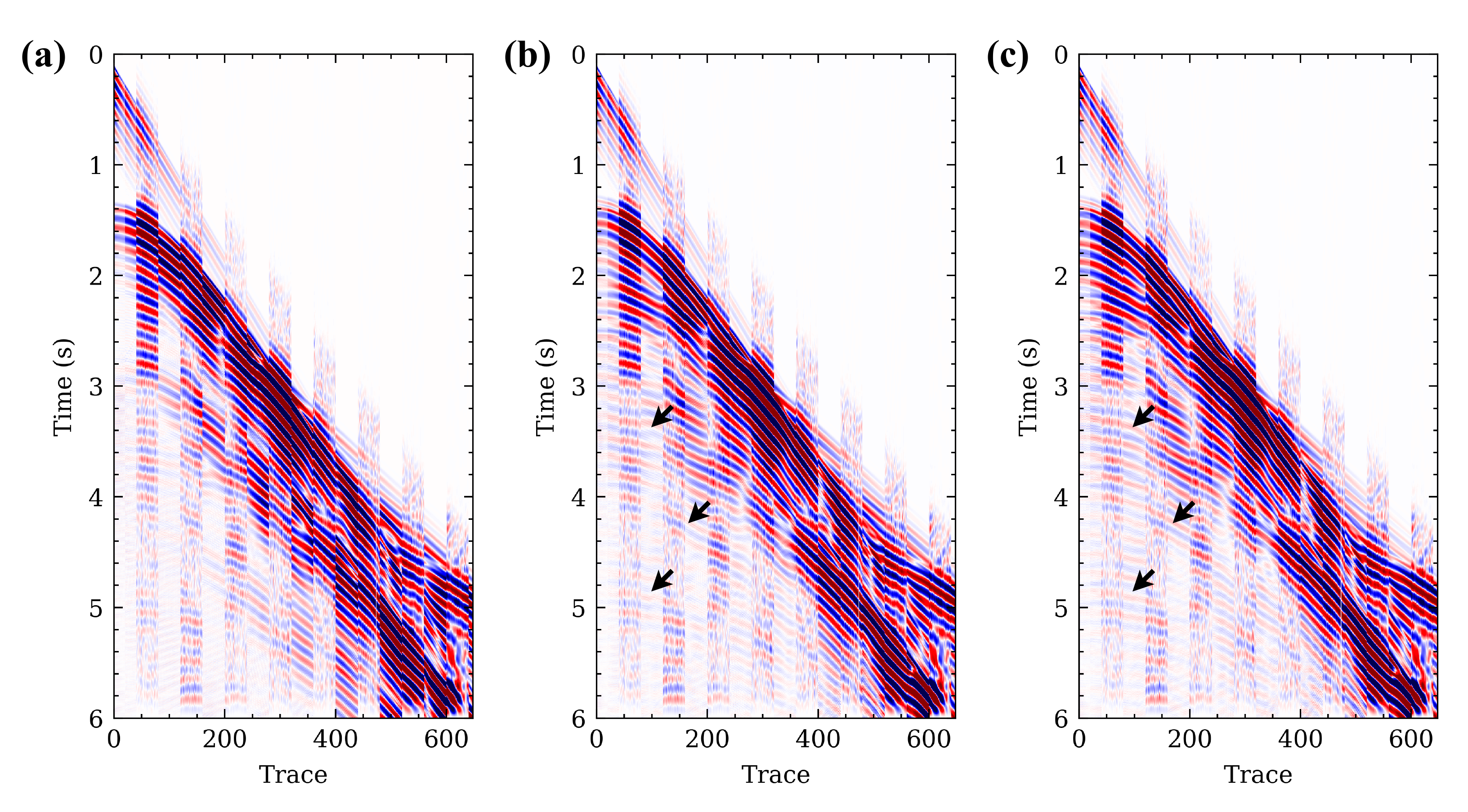}
  \caption{A shot gather in which (a), (b) and (c) display interleaved predicted and observed data using the initial velocity, (Figure~\ref{fieldresult}a), the velocity model from conventional FWI (Figure~\ref{fieldresult}b) and the velocity model of our diffusion regularized FWI (Figure~\ref{fieldresult}c), respectively. 
  We intersperse 40 traces starting from the observed followed by predicted data. }
\label{field_shot_matching}
\end{figure}

To further evaluate the accuracy of the inverted velocity models, we compare their simulated data to the observed ones, as shown in Figure~\ref{field_shot_matching}. 
In general, both simulated data match well with the observed data, especially compared to those from the initial model. 
Nevertheless, the amplitude in the near offset using the proposed method (Figure \ref{field_shot_matching}c) is closer to the
observed data, and seismic events are more continuous, especially for the parts denoted by the arrows.
This is attributed to the sharper interfaces in our inverted model using diffusion FWI.
\begin{figure}[!htb]
  \centering
  \includegraphics[width=0.76\columnwidth]{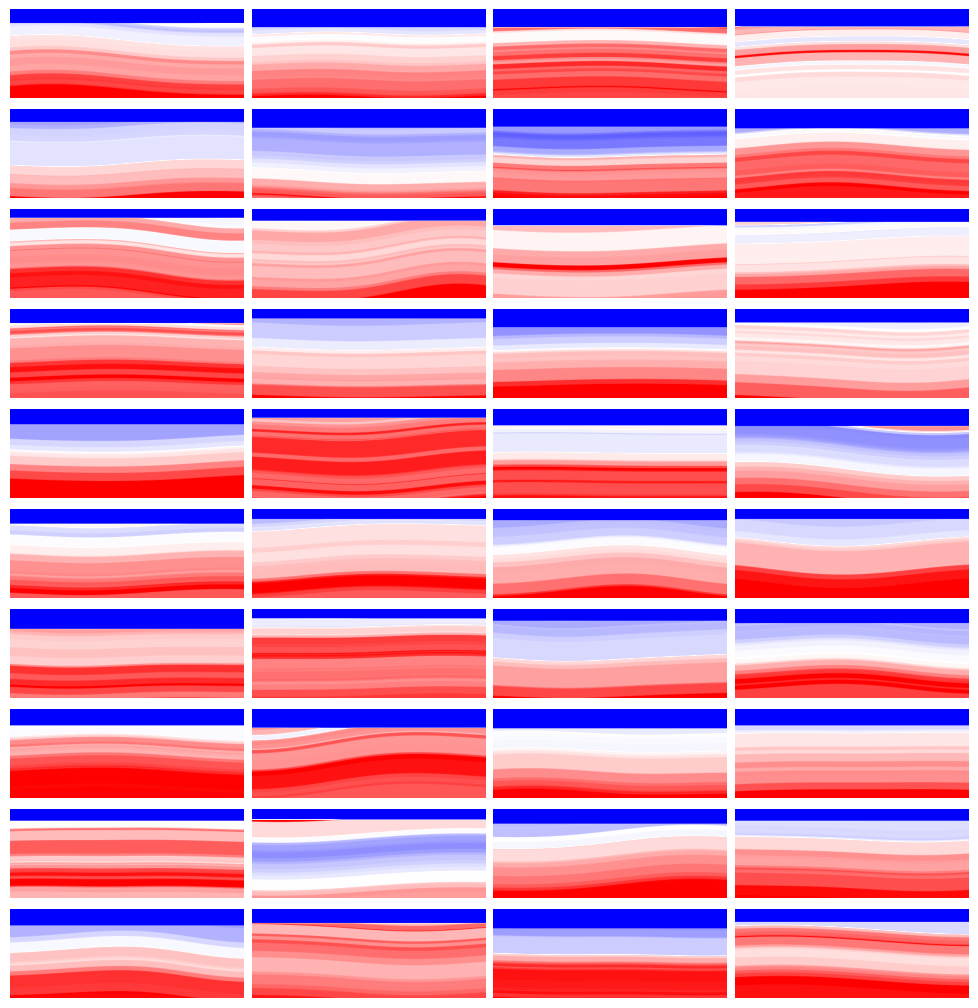}
  \caption{Samples of the generated random velocity models used for training the diffusion model, which were guided by a well in the area and our prior knowledge of the subsurface there.}
  \label{results_field}
\end{figure}

Although the inverted result of diffusion regularized FWI is better than that using the conventional FWI, the results might not be as accurate as we would hope since our prior information stored in the diffusion model is given by the OpenFWI dataset, which does not represent our expectations of the subsurface underneath the observed data. It only provided, as we saw earlier, higher resolution features.
Thus, we repeat the process by retraining the diffusion model with random velocity models guided by the well information and our expectations of the structure in the region, which we expect to be smoother than that given by the OpenFWI dataset.
Specifically, we generate 2000 random velocity models guided by these priors using the exact implementation in \citep{kazei2021mapping,ovcharenko2022multi} and train the diffusion model with this new set. As shown in Figure~\ref{results_field}, the newly generated models are piecewise smoother with less structure.
The training settings are the same as that used in the synthetic examples except for increasing the number of diffusion timesteps to 1000 and the epochs to 400. 
In this way, we could store the priors, e.g., geological priors or the well-log velocity priors in the diffusion model, which is guided by the distribution of the random models used in training. 
Again, due to the relatively poor performance of the DDPM method for high dimensional images, here we also do not directly train the diffusion model with large (296$\times$1000) velocity models. 
In contrast, we train them with velocity models of 152$\times$152 size, which are patches extracted from the original downsampled velocity (with a downsampling factor of 2). 
The calculation of the regular-size velocity models is the same as we described before.

\begin{figure}[t]
  \centering
  \includegraphics[width=0.85\columnwidth]{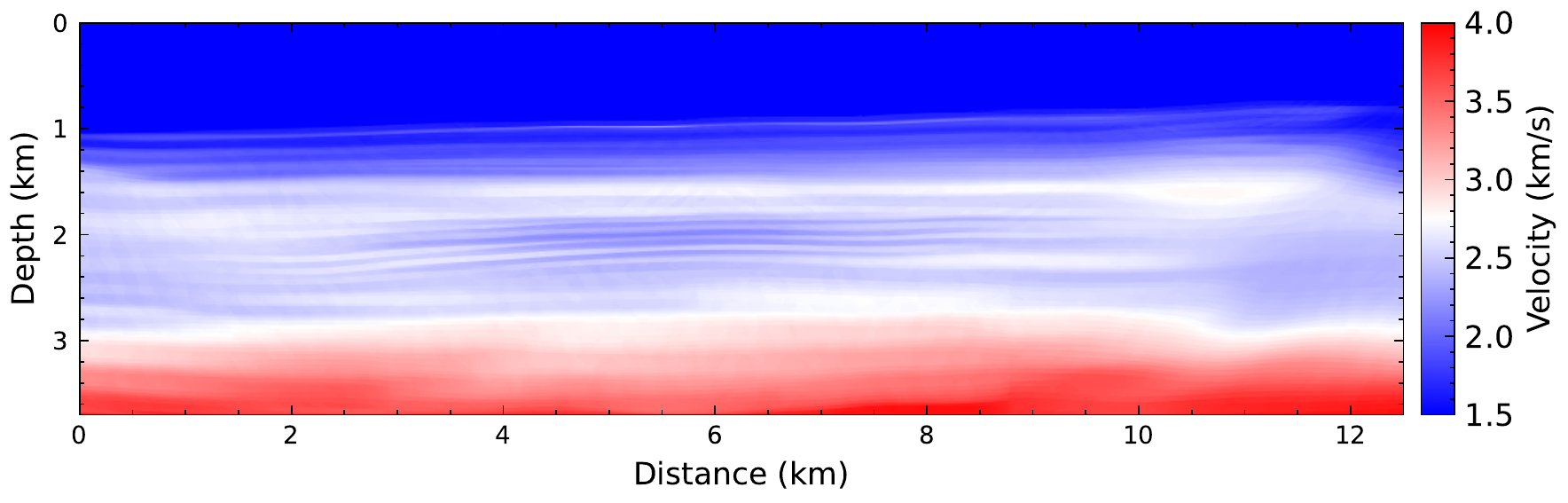}
  \caption{The inverted velocity using the diffusion regularized FWI trained on velocity model samples, some of which are shown in Figure~\ref{results_field}.}
  \label{diffusionfwi_2}
\end{figure}

\begin{figure}[!htb]
  \centering
  \includegraphics[width=0.7\columnwidth]{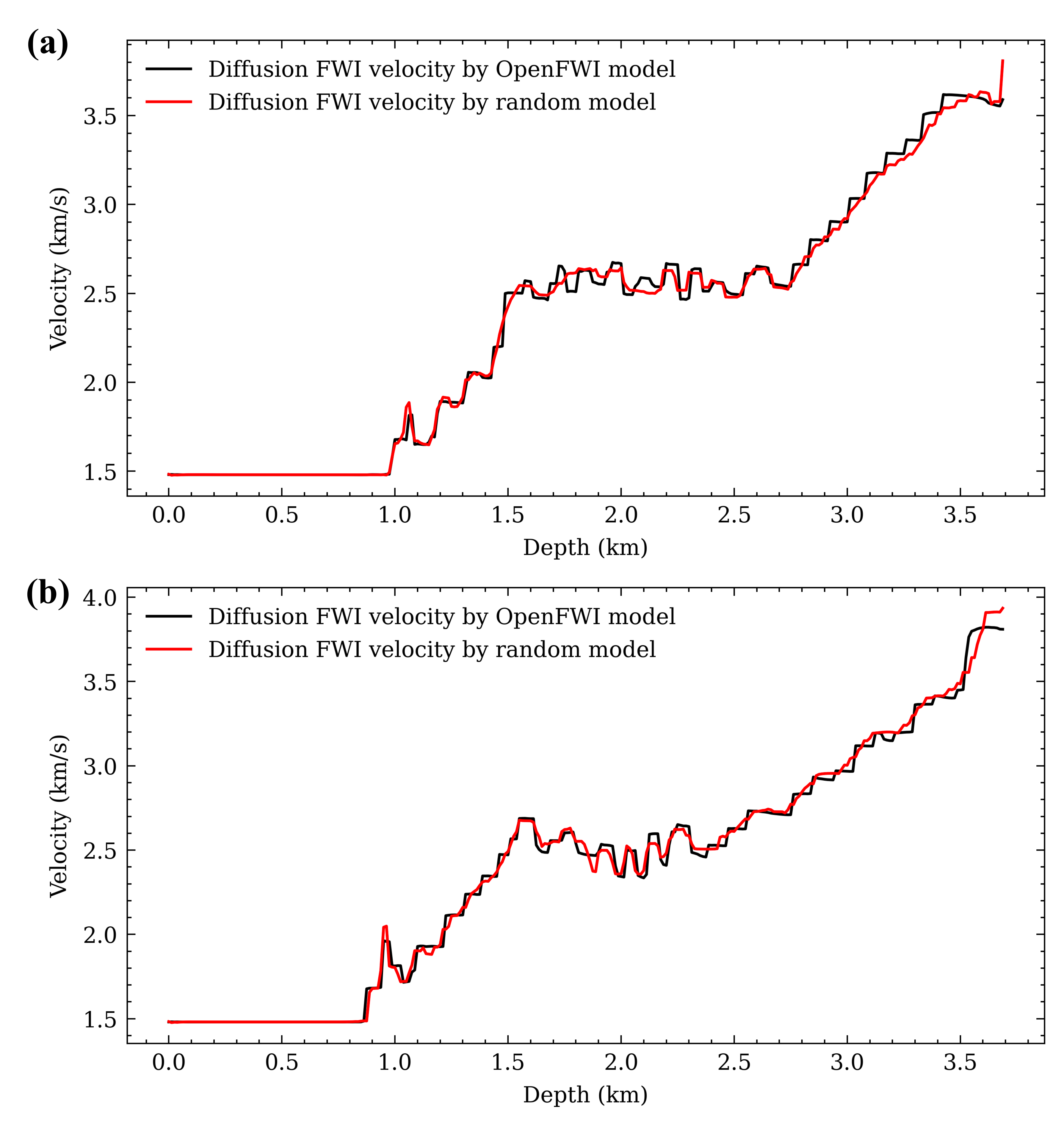}
  \caption{A comparison between vertical profiles at lateral locations 2.5 km (a) and 7.5 km (b),
where the black lines denote the inverted result using the diffusion
regularized FWI based on the pre-trained model using OpenFWI velocity models, and the red lines denote that using the velocity models in Figure~\ref{results_field}.}
  \label{profile2}
\end{figure}

\begin{figure}[!htb]
  \centering
  \includegraphics[width=0.66\columnwidth]{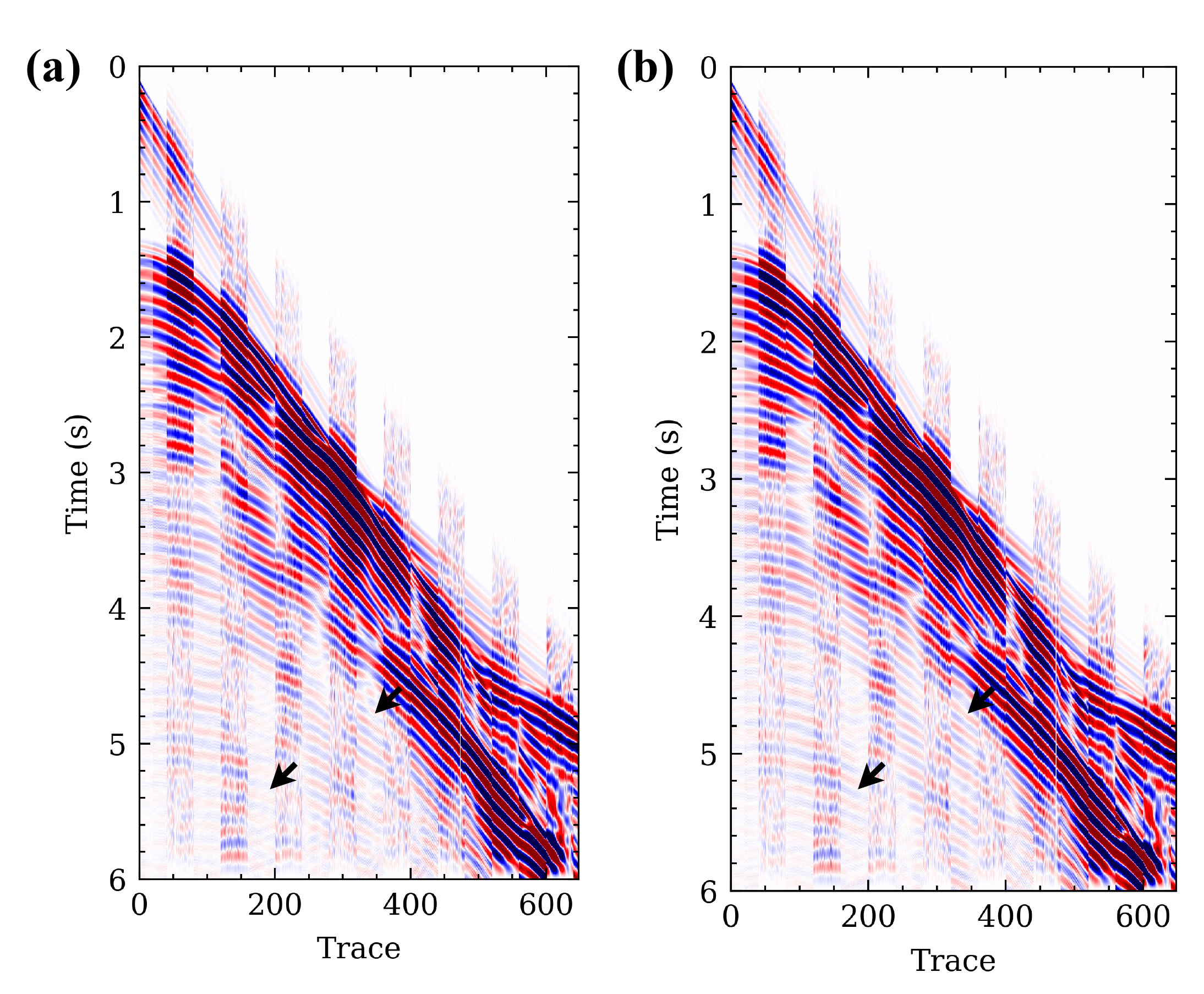}
  \caption{A shot gather in which (a) and (b) display interleaved predicted
and observed data using the velocity model of our diffusion
regularized FWI (Figure~\ref{fieldresult}c) and the velocity model in Figure~\ref{diffusionfwi_2}, respectively. We intersperse 40 traces, starting
from the observed followed by predicted data.}
  \label{shot_gather_2}
\end{figure}

Then we applied the diffusion regularized FWI, and the results are shown in Figure~\ref{diffusionfwi_2}. We can see that the model still maintains the higher resolution feature in the vertical direction, but the lateral continuity is much better than the inverted result built on the pre-trained model with the OpenFWI dataset and the inverted result using the conventional FWI. The artifacts at the right corner, due to the limited illumination, are successfully mitigated. We extract vertical profiles (Figure~\ref{profile2}). With better prior information guidance, the diffusion regularized FWI allows for more reasonable high-wavenumber details. As for the comparison of simulated data using the inverted velocities to the observed data, there are obvious improvements denoted by the arrows in Figure~\ref{shot_gather_2}. This demonstrates that the better the prior information we have, the better the inverted results we obtain.

\section{Discussions}
The proposed method shows good performance in all settings with only a minor additional cost. The computational cost comes from the sampling of the diffusion model, but it is practically instant. On the other hand, the cost of training the diffusion model could be high, but, as mentioned earlier, that can be considered as an overhead cost. Nevertheless, the training process is important to the success of the inversion, so we will discuss possible concerns in this matter.
\subsection{Determining the velocity models used in training}
The quality of the reconstruction of our method highly depends on the distance between the target subsurface model and the distribution of the training dataset.
Taking the field data experiment as an example, although we still obtained a result that is better than conventional FWI using the pre-trained model on OpenFWI dataset, we ended up with less lateral continuity and variations in the bottom right corner as compared to the result using the random velocities in Figure~\ref{results_field}.
The diffusion model trained with improper prior will decrease the ability of the proposed method to mitigate artifacts, e.g., due to the limited illumination, and may lead to unnatural inverted results.
This demonstrates the importance of a proper prior, which should be reasonably close to the inversion target, in training FWI diffusion for accurate velocity estimation. 

As mentioned earlier, we could use various open velocity models for training, e.g., Overthrust model \citep{aminzadeh19963}, Marmousi model \citep{brougois1990marmousi}, and so on. Considering the training cost and the complexity of the diffusion model, we can choose random velocity models, guided by the well log information or geological information \citep{kazei2021mapping}, to train the diffusion model, like what we have done in Section~\ref{sec:fielddata}. 
\subsection{The training of diffusion models}
To properly generate velocity models using the diffusion model, the dimension of the models involved (the depth and width of the network) should be related to the amount of features in the velocity model distribution we need to store. Unlike natural images, velocity models, especially those we generate randomly, do not include the variety, details, and variations seen in natural images. So, we often need smaller dimensional diffusion models to accommodate patches of the velocity model. 
This is actually good as smaller diffusion models are cheaper to train. 
As mentioned earlier, the patches are extracted from the model along the horizontal direction. We do that so all the patches experience the general velocity variation trend with depth. For patches extracted along the vertical direction, the velocity content of a shallow patch will be very different than that for a deeper patch.  
To address this issue of the discrepancy in the distribution of shallow patches compared to deeper ones, as velocity generally increases with depth, we can adapt the depth of the patch as an additional input to the network as in \citep{li2022target}.

\subsection{The resolution component of diffusion models}
For velocity models, we observed that the high-resolution information is often stored in the U-net model within the last few time steps of the reverse diffusion process. The reason for that is the fact that the high-resolution components of an image are often the first to be diffused when we add noise since these features (high wavenumbers) tend to have a low amplitude spectrum in velocity models and even in natural images. This control of the resolution features in the diffusion model-based generator is a unique attribute that is not provided by other generators, and it is especially beneficial for FWI. As a result, we often only utilize the reverse diffusion (generator) time steps that correspond to enhancing the resolution of the velocity model.

\section{Conclusions}
We proposed a new paradigm for regularized FWI using generative diffusion models. 
We use the generative diffusion model to extract the features of velocity
distribution in the training of the diffusion model and tactfully fuse such features into FWI.
Specifically, we pre-train a diffusion model on a prior distribution of velocity models that represent what we expect of the subsurface, probably guided by a well, in a fully unsupervised manner and then adapt it to the seismic observations, incorporating the FWI into the sampling progress of the generative diffusion models. Numerical examples demonstrate that our method can outperform the conventional FWI with only a small additional computational cost. Even in cases of sparse observations or observations with strong noise, the proposed method could still reconstruct a high-quality subsurface model.
A further marine field data test also demonstrates the effectiveness of the proposed method.
With the reverse diffusion generative process, the higher resolution components appear in
the final time steps, which is favorable to FWI.

\section*{Acknowledgment}
We thank KAUST and the DeepWave Consortium sponsors for their support and the SWAG group for the collaborative environment. This work utilized the resources of the Supercomputing Laboratory at KAUST, and we are grateful for that.

\bibliographystyle{plainnat}
\bibliography{diffusionFWI}
\end{document}